\documentclass[aps,prl,twocolumn,superscriptaddress,showpacs]{revtex4}
\usepackage{amsfonts}
\usepackage{amssymb}
\usepackage{amsmath}
\usepackage{nicefrac}
\usepackage{epsfig}
\usepackage{graphicx}
\usepackage{dcolumn}
\usepackage{bm}
\usepackage{flafter}
\usepackage{float}
\usepackage{color}

\begin{document}

\title{Effect of Applied Orthorhombic Lattice Distortion on the Antiferromagnetic Phase of CeAuSb$_2$}
\author{Joonbum Park}
\affiliation{Max Planck Institute for Chemical Physics of Solids, N\"{o}thnitzer Stra{\ss}e 40, 01187 Dresden, Germany}
\affiliation{Max Planck POSTECH Center for Complex Phase Materials, Pohang University of Science and Technology, Pohang 37673, Republic of Korea}
\author{Hideaki Sakai}
\affiliation{Department of Physics, Osaka University, Toyonaka, Osaka 560-0043, Japan}
\affiliation{PRESTO, Japan Science and Technology Agency, Kawaguchi, Saitama 332-0012, Japan}
\author{Onur Erten}
\affiliation{Max Planck Institute for the Physics of Complex Systems, N\"{o}thnitzer Stra{\ss}e 38, 01187 Dresden, Germany}
\author{Andrew P. Mackenzie}
\email{mackenzie@cpfs.mpg.de}
\affiliation{Max Planck Institute for Chemical Physics of Solids, N\"{o}thnitzer Stra{\ss}e 40, 01187 Dresden, Germany}
\affiliation{Scottish Universities Physics Alliance (SUPA), School of Physics and Astronomy, University of St. Andrews,
St. Andrews KY16 9SS, United Kingdom}
\author{Clifford W. Hicks}
\email{hicks@cpfs.mpg.de}
\affiliation{Max Planck Institute for Chemical Physics of Solids, N\"{o}thnitzer Stra{\ss}e 40, 01187 Dresden, Germany}
\date{19 July 2017}

\begin{abstract}
We study the response of the antiferromagnetism of CeAuSb$_2$ to orthorhombic lattice distortion applied through in-plane uniaxial pressure. The response to pressure applied along a $\langle 110
\rangle$ lattice direction shows a first-order transition at zero pressure, which shows that the magnetic order lifts the $(110)/(1\bar{1}0)$ symmetry of the unstressed lattice. Sufficient $\langle
100 \rangle$ pressure appears to rotate the principal axes of the order from $\langle 110 \rangle$ to $\langle 100 \rangle$. At low $\langle 100 \rangle$ pressure, the transition at $T_N$ is weakly
first-order, however it becomes continuous above a threshold $\langle 100 \rangle$ pressure.  We discuss the possibility that this behavior is driven by order parameter fluctuations, with the
restoration of a continuous transition a result of reducing the point-group symmetry of the lattice.
\end{abstract}

\pacs{71.70.Ej, 71.18.+y, 71.20.Nr}

\maketitle

\textbf{I. INTRODUCTION} \\

Transitions in condensed matter systems are defined by their broken symmetries. Electronic orders can sometimes lift the point-group symmetry of their host lattices, for example
two-fold rotationally symmetric order on a tetragonal lattice. This is an intriguing possibility in part because fluctuations can have non-trivial effects on such transitions~\cite{Fernandes12}.
Condensation of a particular order can also obscure strong sub-leading susceptibilities to alternative orders, which one wants to know about to construct a good theory of the processes driving phase
formation~\cite{Kretzschmar13}. As we present in this article, uniaxial pressure can be used to probe both of these possibilities.

We study the heavy-fermion antiferromagnet CeAuSb$_2$, a layered, tetragonal compound with N\'{e}el temperature $T_N=6.5$~K~\cite{Thamizhavel03, Balicas05}. We found in-plane uniaxial pressure to have
a strong effect on the magnetic transition, in ways more pertinent to general questions of how magnetic order and lattice symmetry interact than specifically to heavy-fermion physics. An important
aspect of our work is that the pressure is applied using piezoelectric actuators, allowing in situ tunability. For example, upon ramping the pressure at constant temperature a first-order transition,
with hysteresis, is observed at zero pressure. This transition's existence proves that the magnetic order lifts the point-group symmetry of the lattice.  We also probe a long-standing prediction, that
a transition driven first-order by fluctuations should become continuous when uniaxial pressure selects a preferred direction~\cite{Domany77, KnakJensen79}, with much higher resolution than before.
Finally, there is no requirement to apply high pressures at room temperature, where samples are more susceptible to plastic deformation than at low temperature.

All measurements here were done in zero magnetic field.  Our samples were grown by a self-flux method~\cite{Canfield92, Canfield01}, and have residual resistivity ratios (RRR) $R(\mathrm{300
K})/R(\mathrm{1.5 K})$ between 6 and 9. A shoulder in the resistivity $\rho(T)$ of CeAuSb$_2$ marks the Kondo temperature, $T_K \sim 14$~K~\cite{Seo12}. Therefore at $T_N$ the cerium moments should be
incorporated into the Fermi sea, and there is thermodynamic evidence that they are: The heat capacity has a Fermi liquid form (\textit{i.e.} proportional to $T$) between $T_N$ and $\sim$10~K, and
below $T_N$ shows good entropy balance with a Fermi liquid~\cite{Zhao16}. Recent neutron scattering data suggest that the magnetic order itself is itinerant~\cite{Marcus17}: It was found to be an
incommensurate spin density wave polarised along the $c$ axis, meaning that the polarisation must vary from site to site, which is not expected for local-moment order. The propagation vector is
$(\eta, \eta, 1/2)$, with $\eta$ varying between 0.130 and 0.136 depending on field and field history.

Our apparatus is described in Ref.~\cite{Hicks14_RSI}. Briefly, samples are prepared as beams with high length-to-thickness and length-to-width aspect ratios, and their ends are held in the apparatus
with epoxy, allowing application of both compressive and tensile stresses along their length. A photograph of a mounted sample is shown in Fig.~1(a). The sample is under conditions of uniaxial stress:
the stress is nonzero along the long axis of the sample, and zero along transverse directions. However because the apparatus has a high spring constant relative to that of typical samples, the applied
displacement is the more directly controlled variable than the applied force. For samples such as CeAuSb$_2$, whose elastic moduli have not been reported, longitudinal strain is the controlled
variable and the stress is not known. The displacement applied to the sample and sample mounting epoxy is measured with a capacitive displacement sensor, and the strain in the sample is estimated as
this displacement times 0.8 divided by the exposed length of the sample. The factor of 0.8 is an estimate for the effects of elastic deformation of the epoxy, which allows some relaxation of the
sample strain~\cite{Hicks14_RSI, Hicks14_Science}.

We measure the resistivity along the length of samples, which is strongly affected by the magnetic order~\cite{Thamizhavel03, Balicas05}. The strain-induced change in sample resistance has a geometric
contribution due to the applied change in sample dimensions, typically of magnitude $\Delta R / R \sim 2 \varepsilon$, where $\Delta R$ is the change in resistance and $\varepsilon$ the applied
strain~\cite{Kuo13}. We find $R$ to vary much more strongly with strain than this, and so neglect this geometric effect in all plots and analysis below.  
\\

\begin{figure}[t]
\includegraphics*[width=8.2cm]{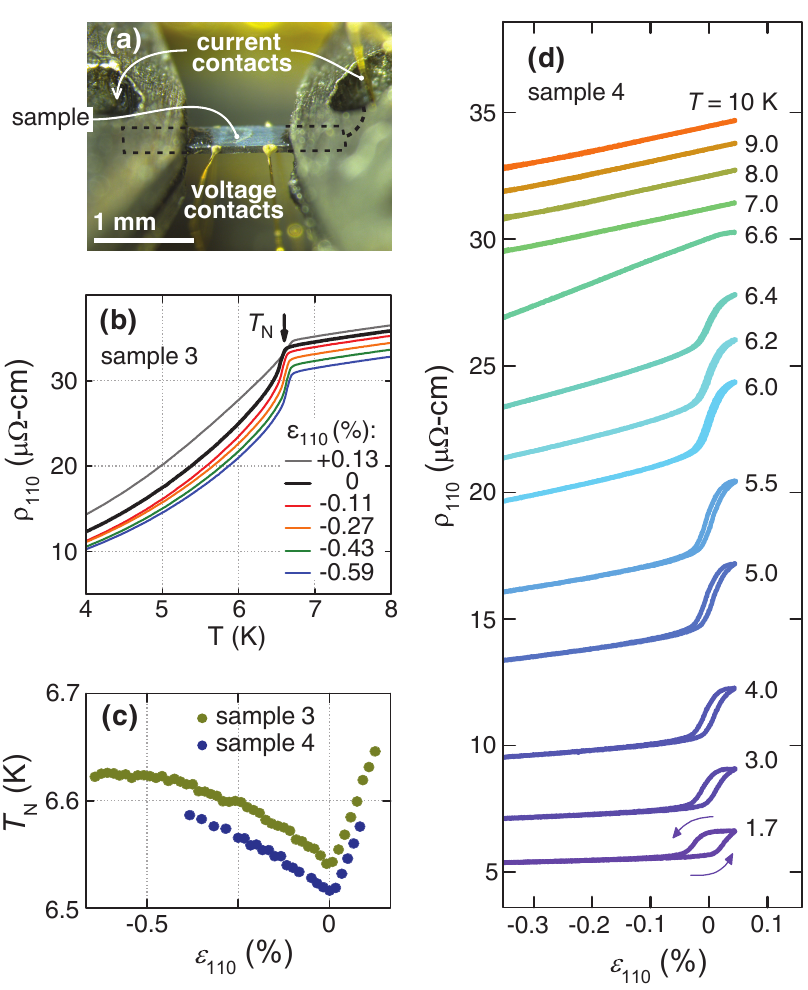}
\caption{\label{Fig.1} (color online) \textbf{Results for $\langle$110$\rangle$ pressure.} (a) Photograph of a mounted sample. (b) Resistivity $\rho_{110}(T)$ along a $\langle 110 \rangle$ lattice
direction at various fixed strains $\varepsilon_{110}$. (c) N\'{e}el temperature $T_N(\varepsilon_{110})$, identified at each strain as the maximum in $d\rho_{110}/dT$. (d)
$\rho_{110}(\varepsilon_{110})$ at various fixed temperatures. Arrows indicate the direction of the strain ramp. In panels (b)-(d), $\varepsilon_{110}=0$ is set to the location of the cusps in
$T_N(\varepsilon_{110})$, seen in panel (c).}
\end{figure}

\textbf{II. RESULTS: $\langle$110$\rangle$ PRESSURE} \\

Five samples were measured under pressure, three cut along a $\langle 100 \rangle$ lattice direction (that is, along the Ce-Ce bond direction), and two along a $\langle 110 \rangle$
direction. Results for $\langle 110 \rangle$ pressure (inducing longitudinal strain $\varepsilon_{110}$) are presented in Fig.~1 panels (b) through (d). Panel (b) shows resistivity versus temperature
at various applied strains. The N\'{e}el transition is clearly seen in each curve, and an immediately apparent result is that the quantitative effect of $\langle 110 \rangle$ pressure on $T_N$ is
small: compression by 0.6\% shifts $T_N$ by only $\sim$0.1~K. 

However, when $T_N$ is plotted against $\varepsilon_{110}$, in panel (c), a sharp cusp in $T_N(\varepsilon_{110})$ becomes apparent. If the cusp is at $\varepsilon_{110} = 0$ it indicates a
two-component order parameter, in which each component lifts the $(110)/(1\bar{1}0)$ symmetry of the lattice. Under this hypothesis, $\langle 110 \rangle$ pressure favors one of these components, and
the favored component switches when the sign of the pressure changes, yielding the sharp change in slope $dT_N/d\varepsilon_{110}$. It is a reasonable hypothesis that the cusp marks
$\varepsilon_{110}=0$.  Firstly, the strain applied to reach the cusp, $\sim 0.1$\%, is compatible with plausible differential thermal contractions between the sample and apparatus frame (which is
made of titanium).  Secondly, samples fractured when tensioned by more than $\sim$0.2\% beyond the cusp, so at that point they were definitely under tension.

If the two components coexist microscopically over some strain range, strain ramps below $T_N$ should show two transitions, corresponding separately to the onset of one and the disappearance of the
other component~\cite{Brodsky17}. If they do not coexist, the ordered state lifts the $(110)/(1\bar{1}0)$ symmetry of the lattice, and a first-order transition, corresponding to reversal of the sign
of the symmetry-breaking, is expected at $\varepsilon_{110}=0$.  Our results, shown in panel (d), show a first-order transition: $\rho(\varepsilon_{110})$ changes in a step-like manner, and there is
clear hysteresis. Within our resolution, it extends up to $T_N$. The neutron data also point to spontaneous symmetry breaking: the observed scattering peaks correspond to incommensurate spin density
wave propagation vectors $\mathbf{q} = (\eta, \eta, \nicefrac{1}{2})$ and $(\eta, -\eta, \nicefrac{1}{2})$, and the absence of peaks corresponding to mixing of these components indicates that they
exist in separate domains~\cite{Marcus17}. Therefore, we conclude firmly that the magnetic order spontaneously lifts the $(110)$/$(1\bar{1}0)$ symmetry of the lattice, and assign the location of the
cusp as $\varepsilon_{110}=0$.

\begin{figure}[t]
\includegraphics*[width=8.2cm]{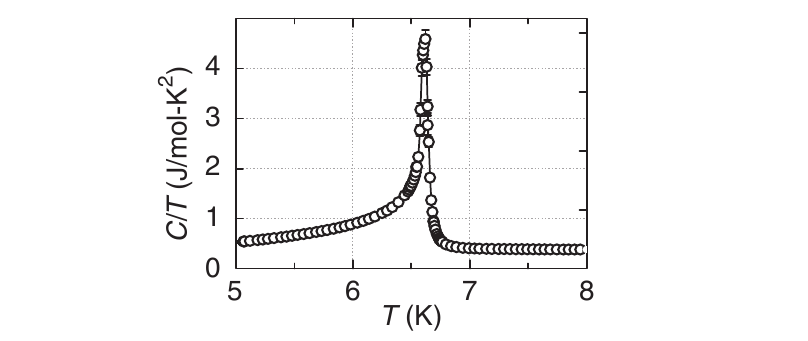}
\caption{\label{Fig.2} Heat capacity of an unpressurized crystal of CeAuSb$_2$, of mass 2.2~mg. The sharp peak is consistent with a first-order transition.}
\end{figure}

The cusp in $T_N(\varepsilon_{110})$ is not symmetric: $|dT_N/d\varepsilon_{110}|$ is smaller on the compressive than on the tension side of the cusp. This is not surprising: uniaxial pressure applies
not only an in-plane orthorhombicity, \textit{i.e.} a nonzero $\varepsilon_{110} - \varepsilon_{1\bar{1}0}$, but also changes to unit cell volume and $c$-axis lattice parameter. Coupling to the latter
two variables will introduce such asymmetry.

Another feature apparent in the data above, also noted in Ref.~\cite{Seo12} and which will be important in discussing results of $\langle 100 \rangle$ pressure, is that the transition at $T_N$ appears
to be weakly first-order. Although we did not resolve hysteresis between increasing- and decreasing-temperature ramps, there is a clear step in $\rho(\varepsilon_{110})$ at $T_N$. For
further evidence, the heat capacity of an unstrained crystal was measured, with the results shown in Fig.~2. The sharp peak in heat capacity at $T_N$ strongly suggests a first-order transition. 
\\

\textbf{III. RESULTS: $\langle$100$\rangle$ PRESSURE} \\

We now turn to results from $\langle 100 \rangle$ pressure. As described in the Introduction, there is strong evidence that the magnetic order of CeAuSb$_2$ is an itinerant order, with the Ce magnetic
moments incorporated into the Fermi sea through the Kondo effect at a temperature well above $T_N$. The Kondo temperature of CeAuSb$_2$ has been shown to be tunable with hydrostatic
pressure~\cite{Seo12}: it increases by a factor of $\sim$2 under 2~GPa. 

We find that uniaxial pressure has a much smaller effect. In Fig.~3 we show the longitudinal resistivity of a sample cut along a $\langle 100 \rangle$ crystal direction at a few applied strains
$\varepsilon_{100}$. The shoulder at $\sim$14~K indicates the Kondo temperature, and is at constant temperature to within 1~K over the range of strains studied. Similarly, $T_K$ of CeRu$_2$Si$_2$ was
also found to have a very weak dependence on in-plane uniaxial pressure~\cite{Saha02}. Therefore, over the range of $\langle 100 \rangle$ strains studied in this paper, we may safely assume that
$T_K$ remains well above $T_N$ and that we are probing an itinerant magnetic order.

\begin{figure}[t]
\includegraphics[width=85mm]{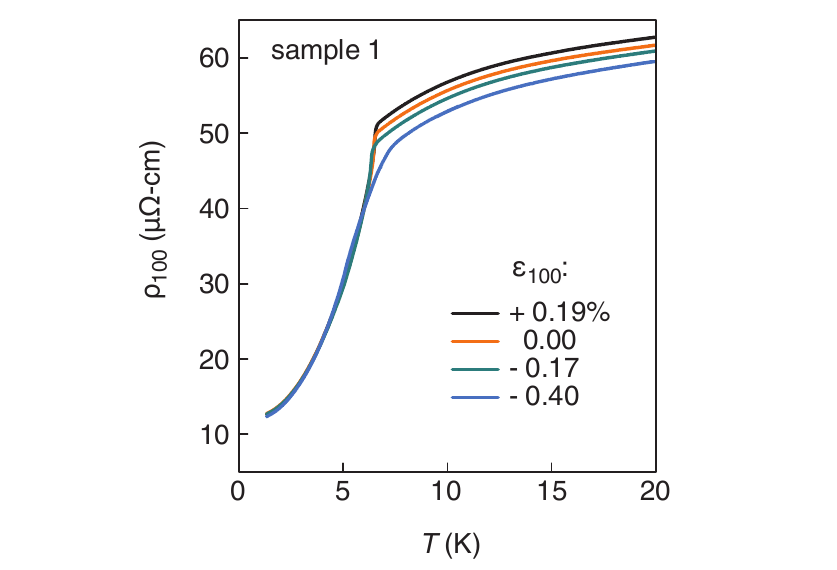}
\caption{\label{Fig.3} Longitudinal resistivity $\rho_{100}(T)$ for sample 1 and various fixed $\varepsilon_{100}$. The shoulder at $\sim$14~K marks the Kondo temperature, which is not strongly affected by strain.}
\end{figure}

$\rho(T)$ at various applied strains for one sample is shown in Fig.~4. To make more clear the first-order-like nature of the transition at low strains, the derivative $d\rho/dT$ is plotted in the
lower panel. In the response to $\langle 100 \rangle$ pressure there is no obvious feature that might be identified with zero strain, so we take zero strain to be at the same applied displacement
where the cusp in $T_N(\varepsilon_{110})$ was observed. Variability in the precise mounting conditions achieved will introduce an error of $\sim$0.1\% on this determination. 

Strains $|\varepsilon_{100}|<0.25$\% do not strongly affect the transition; over this range, compression weakly suppresses $T_N$, and possibly reduces the size of the first-order-like jump in $\rho$.
However at higher compression, $\varepsilon_{100} < -0.25$\%, the first-order jump disappears and the transition splits into two features. The upper feature is a downturn in the slope $d\rho/dT$ and
the lower feature a further downturn; we label their temperatures $T_2$ and $T_1$.  We hypothesize that an equivalent splitting would occur under tensile strain; in the data in Fig.~4, the first-order
jump also shrinks somewhat under tensile pressure. However, attempts to reach this hypothesized splitting in samples 1 and 2 resulted in both samples fracturing, and at a sufficiently low strain,
$\varepsilon_{100} \sim +0.25$\%, that an essentially symmetric strain response is not ruled out.

\begin{figure}[t]
\includegraphics*[width=8.2cm]{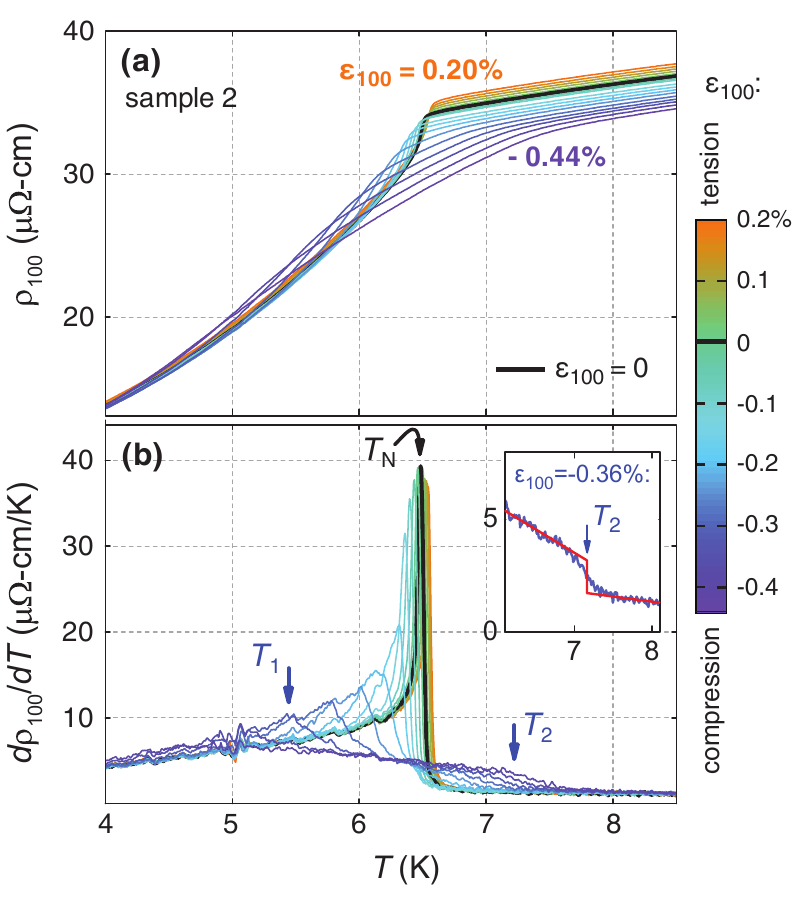}
\caption{\label{Fig.4} (color online) \textbf{Results from $\langle$100$\rangle$ pressure.} (a) Resistivity $\rho_{100}(T)$ along a $\langle 100 \rangle$ lattice direction at various fixed strains
$\varepsilon_{100}$, and (b) the corresponding derivatives $d\rho_{100}/dT$. The temperature $T_1$ is identified as the peak in $d\rho/dT$, and $T_2$ as the step in $d\rho/dT$.}
\end{figure}

\begin{figure}[t]
\includegraphics*[width=8.2cm]{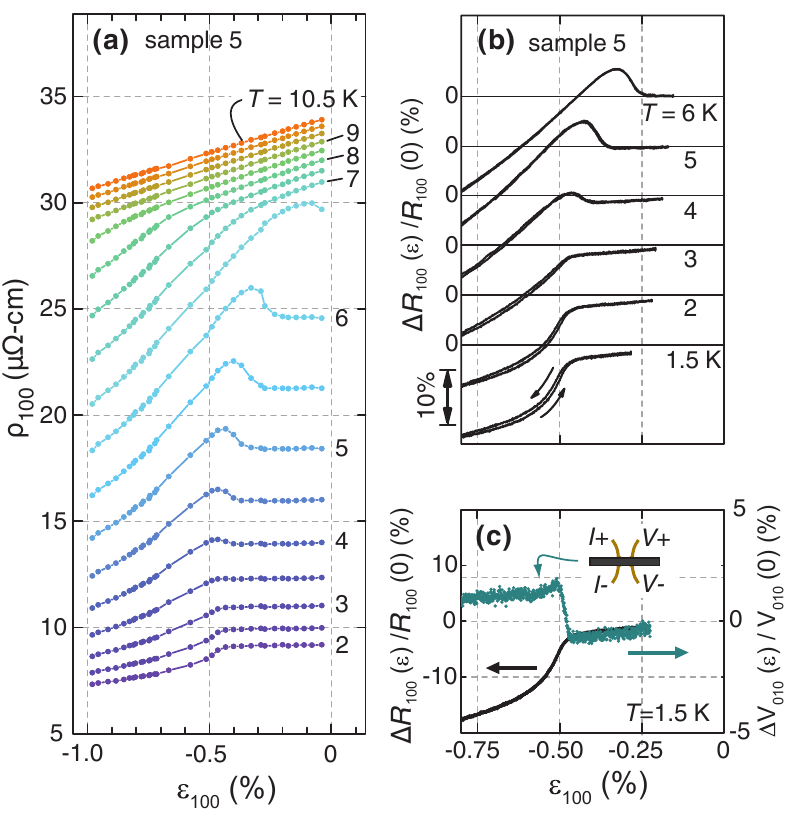}
\caption{\label{Fig.5} (color online) (a) $\rho_{100}(\varepsilon_{100})$ at fixed temperatures on a 0.5~K interval. The data were collected in temperature ramps. (b) Change in
resistance $\Delta R_{100} / R_{100}(\varepsilon_{100}=0)$ during increasing- and decreasing-strain ramps at fixed temperatures. There is hysteresis below $\sim$4~K. (c) For a qualitative measure of
the transverse resistivity $\rho_{010}$, current was applied across the width of one sample, as indicated in the diagram of the contact configuration. The resulting voltage across the sample,
$V_{010}$, is plotted, along with the longitudinal resistance.}
\end{figure}

In Fig.~5 we plot $\rho_{100}$ against $\varepsilon_{100}$ at fixed temperatures. The feature at $T_2$ is visible as a change in slope $d\rho/d\varepsilon$ in the 7.0 to 8.5~K curves. We
assign this feature to be a second-order transition: The slopes $d\rho/d\varepsilon$ and $d\rho/dT$ both change, but no first-order steps in $\rho$ are apparent. 

The feature at $T_1$ is visible in, for example, the 6.0~K curve as a peak in $\rho_{100}$ at $\varepsilon_{100} \sim -0.3$\%. As the temperature is reduced, it moves towards higher compressions and
changes from a peaked into a step-like feature. It is suppressed to below 2~K at $\varepsilon_{100} \sim -0.5$\%. The step-like form at lower temperatures suggests a first-order transition. For
further evidence we performed strain ramps at constant temperature, with the results shown in panel (b). Small hysteresis loops are resolvable at temperatures below $\sim$ 4~K. 

We also probe the transverse resistance, by running current and measuring voltage across the width of the sample, as illustrated in panel (c). In this configuration the current flow is not
homogeneous, so we do not attempt to extract a quantitatively precise transverse resistivity. However the data reveal, as shown in panel (c), that the transverse resistivity changes oppositely to the
longitudinal resistivity across the transition at $\varepsilon_{100} \approx -0.5$\%. Also, the transverse resistivity changes in a sharp, first-order step. We show in Fig.~6 the transverse
resistivity at 1.5~K measured in increasing- and decreasing-strain ramps, which reveal observable hysteresis: It is a first-order transition. Generically, the change in longitudinal resistivity should
be as sharp as the change in transverse resistivity, and a possible reason that in the data it is not is that the transverse configuration probes, effectively, a smaller volume of the sample.

The first-order transition does not appear to extend up to $T_N$. The hysteresis disappears, and the form of $\rho(\varepsilon_{100})$ changes from step-like to peaked at $T \sim 3.5$~K. The peaked
form of $\rho(\varepsilon_{100})$ at higher temperatures may be a result of critical fluctuations above the endpoint of the first-order transition.
\\

\textbf{IV. DISCUSSION AND CONCLUSION} \\

 As explained above, we conclude in agreement with neutron data~\cite{Marcus17} that the magnetic order of unstressed CeAuSb$_2$ spontaneously lifts the $(110)/(1\bar{1}0)$
symmetry of the lattice. The V-shaped form of $T_N(\varepsilon_{110})$ indicates two anisotropic order parameter components, with $\langle 110 \rangle$ principal axes, and the first-order transition
across $\varepsilon_{110}$ shows that they do not co-exist microscopically. We note that Sr$_3$Ru$_2$O$_7$ provides an alternative example, of anisotropic order parameter components that do co-exist
microscopically over a non-zero range of applied lattice orthorhombicity~\cite{Brodsky17}.

\begin{figure}[t]
\includegraphics[width=85mm]{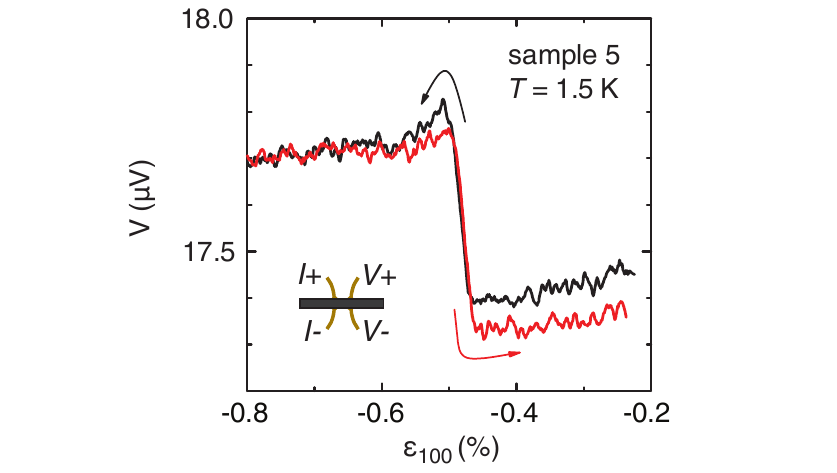}
\caption{\label{Fig.6}Transverse response of the sample across the transition at $\varepsilon_{100} \approx -0.5$\%, from increasing- and decreasing-strain ramps.}
\end{figure}

However, although the spontaneous symmetry breaking is with $\langle 110 \rangle$ principal axes, $\langle 100 \rangle$ pressure has a quantitatively much stronger effect than $\langle 110 \rangle$
pressure.  We summarize our $\langle 100 \rangle$ pressure data with the phase diagram in Fig.~7. It appears very likely that strong $\langle 100 \rangle$ pressure changes the principal axes of the
order from $\langle 110 \rangle$ to $\langle 100 \rangle$, in other words that unstressed CeAuSb$_2$ has a sub-leading susceptibility to a $\langle 100 \rangle$ order which becomes dominant with
sufficient applied $(100)/(010)$ orthorhombicity. The simplest example to imagine is that the in-plane propagation vector of the spin density wave rotates from $(\eta, \pm \eta)$ to $(\eta_2, 0)$ or
$(0, \eta_2)$, with $\eta_2$ in general not equal to $\eta$. A first piece of evidence for rotation of the principal axes is the first-order transition at $\varepsilon_{100} \approx -0.5$\%:
Electronic orders generally pin to high-symmetry directions of the host lattice, so rotation between $\langle 110 \rangle$ and $\langle 100 \rangle$ principal axes should, in general, be
discontinuous. A second is the strong linear dependence of transition temperature $T_2$ on $\langle 100 \rangle$ pressure, in other words on applied $(100)/(010)$ orthorhombicity. In principle, the
linear dependence could also be due to coupling to unit cell volume and/or interplane spacing, parameters that also vary linearly with applied $\langle 100 \rangle$ pressure. However unless the
mechanical properties of CeAuSb$_2$ are extraordinarily anisotropic $\langle 110 \rangle$ pressure will yield similar changes to these parameters, and yet had much less effect on the magnetic
transition. 

In the density functional theory calculations reported in Ref.~\cite{Marcus17}, nesting vectors parallel to $\langle 100 \rangle$ directions as well as $\langle 110 \rangle$ directions were found,
so CeAuSb$_2$ may well have a strong sub-leading susceptibility to a $\langle 100 \rangle$ spin density wave. A more complicated textured order such as the field-induced ``woven'' order proposed in
Ref.~\cite{Marcus17}, which has $\langle 100 \rangle$ principal axes, is also in principle a possibility, however this order was proposed as a way to accommodate both strong nesting and strong
field-induced polarization, and at zero field a straightforward spin density wave seems generically more likely.

The phase diagrams against $\langle 100 \rangle$ and $\langle 110 \rangle$ pressure are our main results. In the remainder of this article we discuss a different topic, the possibility that the
N\'{e}el transition, for $|\varepsilon_{100}| < 0.25$\%, is driven first-order by competing fluctuations, and that uniaxial pressure restores a continuous transition by selecting a preferred direction
and eliminating the competition.

\begin{figure}[t]
\includegraphics[width=8.2cm]{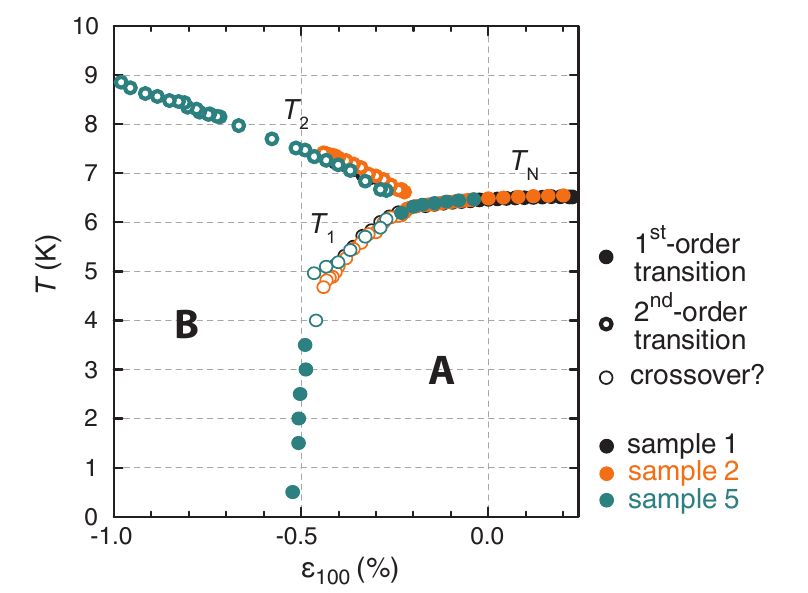}
\caption{\label{Fig.7} (color online) (a) $T$ - $\varepsilon_{100}$ phase diagram. $T_N$ is taken as the maximum in $d\rho/dT$. $T_2$ is taken as the midpoint of the step in $d\rho/dT$, as shown in
Fig.~4(b). $T_1$ is taken at higher temperatures as the peak in $d\rho/dT$ [see Fig.~4(b)], and at lower temperatures as the approximate midpoint of the step in $\rho(\varepsilon_{100})$ [see
Fig.~5(a)]. The in-plane propagation vector is $(\eta, \pm \eta)$ for phase A, and we propose that it is $(\eta_2, 0)$ or $(0, \eta_2)$ for phase B.} 
\end{figure}

We can rule out an alternative explanation for the first-order transition, strong magnetoelastic coupling. If the gain in magnetic condensation energy from a given lattice distortion exceeds its
elastic energy cost, then the transition becomes first-order~\cite{Bean62}. The strain dependence of $T_N$ may be written as $T_N = T_{N,0}(1 + \beta \varepsilon)$, with $\beta$ a coupling constant
and $\varepsilon$ a strain associated with the mode of deformation most strongly favoring the ordered phase. The heat capacity of a material is $C = -T \frac{\partial^2F}{\partial T^2}$, where $F$ is
the free energy. For $T$ close to and below a second-order transition at $T_N$, this expression may be integrated:
\[
\Delta F = \frac{\Delta C}{2 T_N} (T_N-T)^2,
\]
where $\Delta C$ and $\Delta F$ are the change in heat capacity and free energy due to the magnetic order. $\Delta C$, from the data in Fig.~2, is $\sim 1 \cdot 10^5$~J/m$^3$-K. The elastic energy
cost of lattice deformation is $\Delta F = (E/2)\varepsilon^2$, where $E$ is the elastic modulus associated with strain $\varepsilon$. The elastic compliance drives the transition first-order if the
gain in condensation energy exceeds the elastic energy cost, \textit{i.e.} if $E - \Delta C T_{N,0}\beta^2 < 0$.

Although the elastic moduli of CeAuSb$_2$ have not been measured, we may take $E \sim 100$~GPa, a typical Young's modulus for metals, as an order-of-magnitude estimate. Therefore, a first-order
transition is expected if $\beta$ exceeds $\sim 400$. With uniaxial pressure, $dT_N/d\varepsilon$ is not nearly so large: The steepest $\varepsilon \rightarrow 0$ strain dependence is obtained with
tensile $\langle 110 \rangle$ pressure, for which $|dT_N/d\varepsilon| \approx 94$~K, yielding $\beta \approx 14$. We also tested the effect of biaxial pressure, by epoxying thin samples of CeAuSb$_2$
to, respectively, titanium and aluminum plates and using the differential thermal contraction to apply biaxial pressure. The differential thermal contraction between these materials at $T \rightarrow
0$ is 0.25\%, and the observed difference in $T_N$ was 0.04~K, yielding $\beta \sim 2.5$. This measured value of $\beta$ might be suppressed by plastic deformation of the epoxy during the initial
stages of the cool-down, which would relax some of the thermal stress, however it is orders of magnitude too low to drive a first-order transition.

Instead, we propose that the transition is driven first-order by fluctuations. The magnetic order in CeAuSb$_2$ persists even if the RRR is below three~\cite{Thamizhavel03}, indicating a robust order
with a short range of interaction, which favors stronger fluctuation effects~\cite{Ginzburg60, Binder87}. Competition between fluctuations in the disordered state can drive a transition first-order;
in theoretical studies of magnetic helices in MnSi~\cite{Bak80}, density wave order in layered cuprates~\cite{DePrato06, Millis10}, and general multi-component orders~\cite{Domany77}, a continuous
transition is predicted when all possible components of the order can condense simultaneously. However there is competition in CeAuSb$_2$, where condensation of \textit{e.g.} $(\eta, \eta,
\nicefrac{1}{2})$ prevents condensation of $(\eta, -\eta, \nicefrac{1}{2})$ order. Ref.~\cite{Domany77} provides a more precise criterion for fluctuation-driven first-order transitions. Ordered phases
were studied with a fourth-order, multi-component Ginzburg-Landau Hamiltonian, in which competition between the components is set by the biquadratic terms (eq.~2.1 of that paper). The Hamiltonian was
constructed so that either one or all of the components could condense: the coefficients of the biquadratic terms were set equal, and if this coefficient is below a threshold then the components may
co-exist in mean-field theory, and if above condensation of one precludes condensation of all others. It was found that if the number of components $n$ is $\geq 4$, a first-order transition is
expected as soon as the biquadratic coefficient exceeds this threshold, \textit{i.e.} as soon as only one components condenses in mean-field theory. (For $n<4$, stronger competition is required to
get a first-order transition.) $n$ is at least 4 in CeAuSb$_2$: there are two possible density wave orientations, and being incommensurate there are phase and amplitude degrees of freedom for each.

Restoring a continuous transition is predicted to require symmetry-breaking fields exceeding a noninfinitesimal threshold strength~\cite{Domany77, Millis10, KnakJensen79}. The clearest experimental
demonstration is on the antiferromagnetic transition of MnO. It is first-order, but becomes continuous under uniaxial stress~\cite{Bloch75}, a result explained through the effect of reduced point-group
symmetry on fluctuations~\cite{Brazovskii75, Bak76}. However this demonstration is over forty years old and piezoelectric-based pressure apparatus offers much better resolution. In CeAuSb$_2$,
$\langle 100 \rangle$ pressure appears to restore a continuous transition by rotating the principal axes to $\langle 100 \rangle$ and selecting a preferred direction between $(100)$ and $(010)$.
Strong $\langle 110 \rangle$ pressure should also restore a continuous transition, by selecting between the $(110)$ and $(1\bar{1}0)$ directions, however the weak coupling between the electronic
system and $\langle 110 \rangle$ lattice deformation means that this may occur at a pressure beyond what we were able to apply. 

We acknowledge the financial support of the Max Planck Society. JP acknowledges the financial support of the Study for Nano Scale Optomaterials and Complex Phase Materials (No. 2016K1A4A4A01922028)
through NRF funded by MSIP of Korea. HS acknowledges the financial support of PRESTO, JST (No. JPMJPR16R2) and Grant-in-Aid for Young Scientists (No. 16H06015). We thank Collin Broholm and Erez Berg
for useful discussions, and Paul Canfield and Veronika Fritsch for their assistance with sample growth. Raw data for all figures in this paper are available at \textit{to be determined}.

\end{document}